\documentclass{jfm}
\usepackage{graphicx}
\usepackage{epstopdf, epsfig}
\usepackage{url}
\usepackage{amsmath,bm}
\usepackage{amssymb}
\usepackage{todonotes}
\usepackage{subcaption}

\newdimen\figrasterwd
\figrasterwd\textwidth

\newcommand {\Wi}{W\!i}

\shorttitle{Linear stability analysis of purely elastic travelling wave solutions}
\shortauthor{Martin Lellep, Moritz Linkmann, and Alexander Morozov}

\title{Linear stability analysis of purely elastic travelling wave solutions in pressure driven channel flows}

\author{Martin Lellep\aff{1},
  Moritz Linkmann\aff{2},
 \and Alexander Morozov\aff{1}\corresp{\email{alexander.morozov@ed.ac.uk}}}

\affiliation{
\aff{1} SUPA, School of Physics and Astronomy, The University of Edinburgh, James Clerk Maxwell Building, Peter Guthrie Tait Road, Edinburgh, EH9 3FD, United Kingdom
\aff{2}School of Mathematics and Maxwell Institute for Mathematical Sciences, \\ University of Edinburgh, Edinburgh, EH9 3FD, United Kingdom
}

\begin{document}

\maketitle

\begin{abstract}
Recent studies of pressure-driven flows of dilute polymer solutions in straight channels demonstrated the existence of two-dimensional coherent structures that are disconnected from the laminar state and appear through a sub-critical bifurcation from infinity. These travelling-wave solutions were suggested to organise the phase-space dynamics of purely elastic and elasto-inertial chaotic channel flows. Here, we consider a wide range of parameters, covering the purely-elastic and elasto-inertial cases, and demonstrate that the two-dimensional travelling-wave solutions are unstable when embedded in sufficiently wide three-dimensional domains. Our work demonstrates that studies of purely elastic and elasto-inertial turbulence in straight channels require three-dimensional simulations, and no reliable conclusions can be drawn from studying strictly two-dimensional channel flows.
\end{abstract}

\begin{keywords}
% nothing to do here
\end{keywords}

\section{Introduction}

The past few years have seen a growing interest in understanding of pressure-driven channel flows of dilute polymer solutions \citep{Datta2022,Sanchez2022}. The state space of such flows is spanned by three dimensionless parameters: $Re$, the Reynolds number that measures the relative importance of inertia compared to viscous dissipation; $\Wi$, the Weissenberg number that measures the strength of polymer-induced viscoelastic stresses; $\beta$, the solvent to the total solution viscosity ratio that is related to the polymer concentration. The rich variety of flow states associated with different parts of the $(Re,\Wi,\beta)$ space ranges from drag-reduction and elasto-inertial turbulence \citep{Dubief2023} to purely elastic instabilities and turbulence \citep{Steinberg2021}, making it a subject of interest for turbulence and flow-control researchers, polymer rheologists, and physicists studying flows of complex fluids \citep{Datta2022,Sanchez2022}. 

The most significant recent advances in this area correspond to flows with small to negligible amounts of inertia. Such flows were believed to be laminar and exhibit no instabilities, until several  studies challenged this view. While early results on pressure-driven channel flows of model dilute polymer solutions indicated their linear stability \citep{Gorodtsov1967,Wilson1999},  \cite{Morozov2005,Morozov2007,Morozov2019} proposed a sub-critical bifurcation-from-infinity as a non-linear destabilisation mechanism, similar to Newtonian wall-bounded shear flows \citep{Eckhardt2018,Graham2021}. The existence of such a sub-critical bifurcation was demonstrated experimentally by Arratia and co-workers \citep{Pan2013,Qin2017,Qin2019}, while the more recent experiments by Steinberg and co-workers \citep{jha2020preprint, Jha2021, shnapp2021preprint, li2022preprint} provide insight into the spatial structure of the ensuing instability. 

\begin{figure}
\begin{center}
\includegraphics[width=0.7\textwidth]{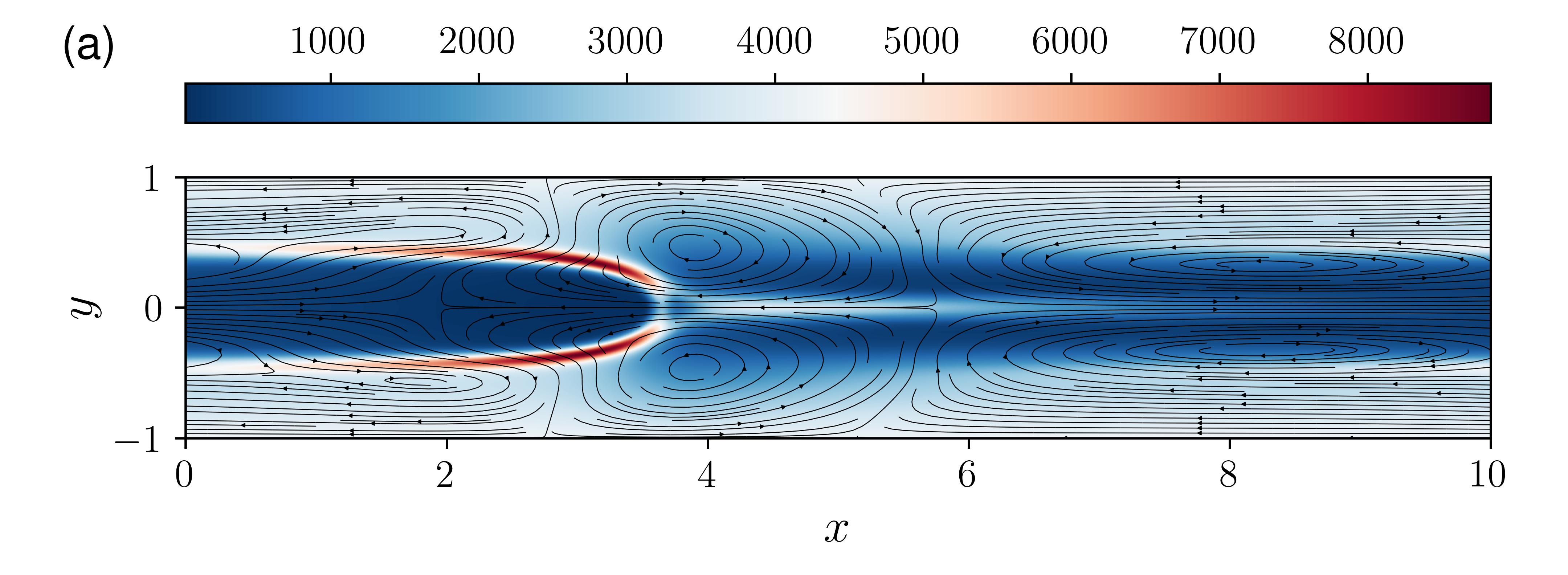} \\ 
\includegraphics[width=0.7\textwidth]{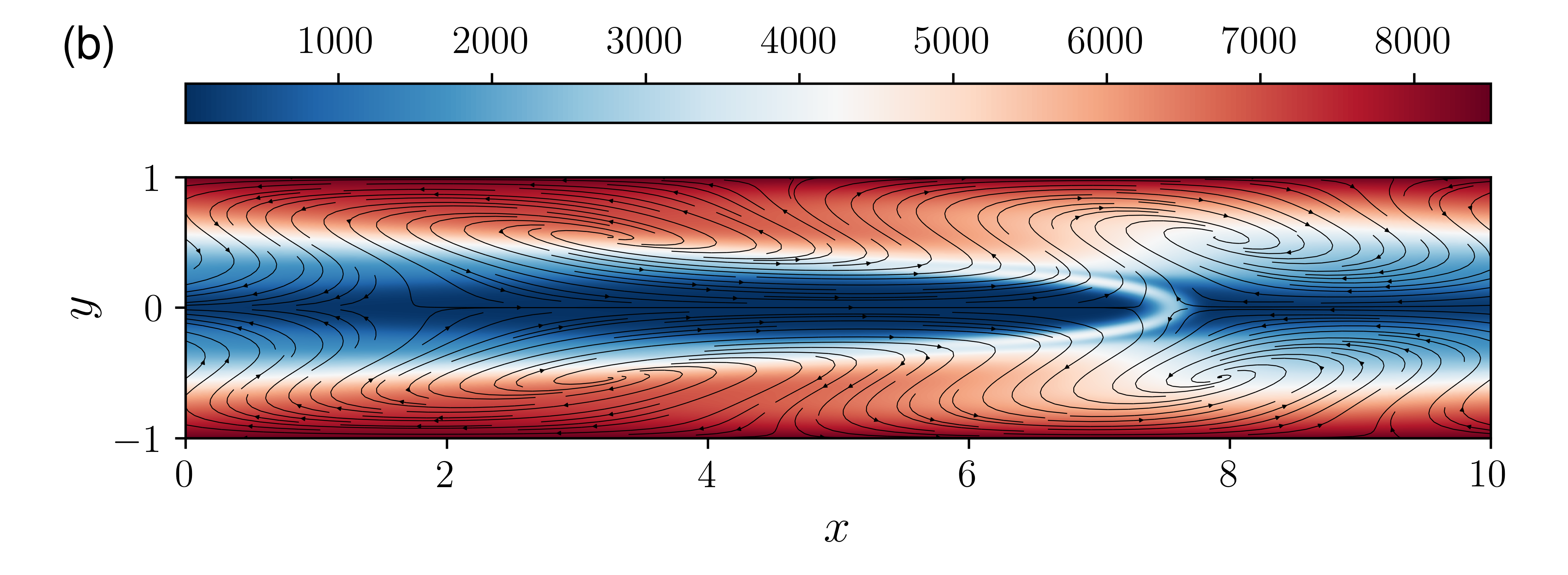} \\ 
\includegraphics[width=0.7\textwidth]{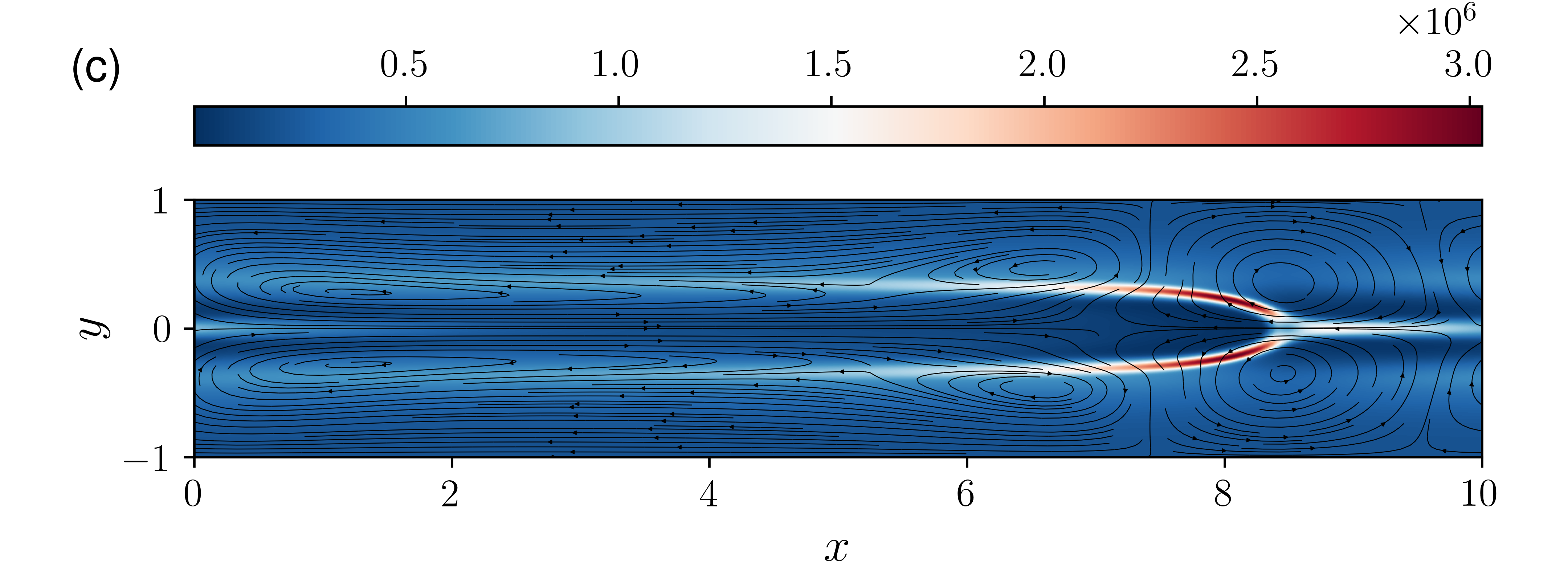}
\end{center}
\caption{Examples of two-dimensional narwhal travelling-wave solutions used as the based state in our linear stability analysis. (a) A purely elastic state with $(\Wi, \beta, Re)$ = $(100, 0.8, 0.01)$ and $L_x=10$, studied in \cite{Morozov2022}. (b) An elasto-inertial state with $(\Wi, \beta, Re)$ = $(45, 0.997, 90)$ and $L_x = 2\pi/2.18$, motivated by the work of \cite{Page2020}. (c) A purely elastic state connected to the centre-mode instability found by \cite{Khalid2021} with $(\Wi, \beta, Re)$ = $(1700, 0.997, 0.01)$ and $L_x = 2\pi/0.75$. The colours indicate values of ${\rm tr}(\bm{c})$ and solid lines show the streamlines of the velocity deviation from the streamwise-averaged flow profile. The mean flow is from left to right along the $x$-direction.
}
  \label{fig:narwhals}
\end{figure}

In addition to these developments, Shankar and co-workers re-visited the linear stability analysis of pressure-driven channel flows of model dilute polymer solutions, and identified a novel linear instability mechanism associated with the centre-line mode \citep{Chaudhary2019, Khalid2021, Khalid2021a}. Although this linear instability exists in a narrow range of parameters, $1-\beta\ll 1$ and $\Wi\gg1$, which can be difficult to simultaneously realise experimentally, the non-linear structures that stem from this instability appear to be crucial to the understanding of elasto-inertial and purely elastic channel flows. \cite{Page2020} and \cite{buza2021} showed that for sufficiently large values of $Re$, the linear instability discussed above leads to a two-dimensional sub-critical bifurcation that extends significantly outside of the $(Re,\Wi)$ region where the linear instability exists. \cite{Morozov2022} performed direct numerical simulations of two-dimensional channel flow at negligible inertia, $Re\ll 1$, for a wide range of $\beta$ and $\Wi$, and reported the existence of purely elastic non-linear travelling wave solutions. These solutions, that we will refer to as the \emph{narwhals}\footnote{Previous work referred to similar structures as \emph{arrowheads} \citep{Page2020,buza2021,Dubief2022}. We do not appreciate the association of that term with warfare, and, instead, prefer to call them \emph{narwhals}, for the obvious likeness of their polymer stress distribution (see Fig.\ref{fig:narwhals}, for example). We are grateful to Prof. Becca Thomases (Smith College) for suggesting the name.}, appear through a bifurcation from infinity that ties their existence with the original proposal by \cite{Morozov2005,Morozov2007,Morozov2019}. Their obvious centre-mode character and their visual resemblance to the structures reported in the elasto-inertial regime \citep{Page2020,buza2021,Dubief2022}, most likely imply that these structures belong to the same solution family that is ultimately connected to the centre-mode linear instability found in a rather special part of the $(\beta,\Wi,Re)$ parameter space.

While the narwhal solutions share many of the features observed in experiments \citep{Pan2013,Qin2017,Qin2019,jha2020preprint, Jha2021, shnapp2021preprint, li2022preprint}, they obviously do not represent chaotic flows; indeed, in two-dimensional channel flows, they are steady travelling wave solutions. To be dynamically relevant, these structures are expected to lose their stability when embedded in three-dimensional domains, as is indeed the case with Newtonian Tollmien-Schlichting waves \citep{Orszag1983}. The purpose of the current work is to study the three-dimensional linear stability of the two-dimensional narwhal solutions discussed above. We report that in all cases studied here for a wide range of $\beta$, $\Wi$, and $Re$, the two-dimensional narwhal solutions lose their stability when embedded in a three-dimensional domain. We characterise this instability and discuss its implications for the mechanism of elastic and elasto-inertial turbulence.

\section{Equations of motion and numerical details}
\label{sec:numerics}

% DNS & model
We consider three-dimensional pressure-driven channel flow and select a Cartesian coordinate system with $(x,y,z)$ being along the flow (streamwise), gradient, and spanwise directions, correspondingly. The dimensionless equations of motion for the polymeric fluid are given by the Phan-Thien-Tanner (PTT) constitutive model \citep{PhanThien1977}, coupled to the incompressible Navier-Stokes equation
\begin{align}
    & \partial_t \bm{c}+\bm{u}\cdot\bm{\nabla}\bm{c}-\bm{\nabla u}^\dagger\cdot\bm{c}-\bm{c}\cdot\bm{\nabla u} = -\frac{\bm{c}-\bm{1}}{\Wi}(1-3\epsilon+\epsilon~ {\rm tr}(\bm{c})) + \kappa \Delta \bm{c}, \label{eq:PTT.constitutive} \\
    & \partial_t \bm{u} + \bm{u}\cdot\bm{\nabla} \bm{u} = -\bm{\nabla}p + \frac{\beta}{Re}\Delta \bm{u} + \frac{1-\beta}{Re \Wi} \bm{\nabla}\cdot \bm{c}+\frac{2}{Re}\bm{e}_x, \label{eq:PTT.momentum} \\
    & \bm{\nabla}\cdot\bm{u} = 0 \label{eq:PTT.incompressibility},
\end{align}
where the dagger denotes the transpose, $\bm{1}$ is the identity matrix, $\rm tr$ denotes the trace, $\bm{e}_x$ is the unit vector in the $x$-direction,  $\bm{u}$ is the fluid velocity,  $\bm{c}$ is the polymer conformation tensor, and $p$ is the pressure. The equations are rendered dimensionless by using $d$, $\mathcal{U}$, $d/\mathcal{U}$, $\eta_p \mathcal{U} / d$, and $(\eta_s+\eta_p)\mathcal{U}/d$, as the scales for length, velocity, time, stress, and pressure, respectively. Here, $d$ is the channel half width, $\eta_s$ and $\eta_p$ are the solvent and polymeric contributions to the viscosity, and $\mathcal{U}$ is the center-line value of the laminar fluid velocity of a Newtonian fluid with the viscosity $\eta_s+\eta_p$ at the same value of the applied pressure gradient. As discussed above, the state diagram of this flow is spanned by the viscosity ratio $\beta=\eta_s/(\eta_s+\eta_p)$, the Reynolds number $Re=\rho\mathcal{U} d/(\eta_s+\eta_p)$, and the Weissenberg number $\Wi= \lambda \mathcal{U} / d$, where $\rho$ is the density of the fluid and $\lambda$ is the Maxwell relaxation time of polymer molecules. Finally, the constant $\epsilon$ controls the degree of shear-thinning of the model PTT fluid, and $\kappa$ is a dimensionless polymer diffusivity. The stress-diffusion term in Eq.~\eqref{eq:PTT.constitutive} is added to ensure that the conformation tensor $\bm{c}$ remains strictly positive-definite at all times throughout out work. We note that while some previous studies employed unrealistically large values of $\kappa$ for this purpose, the value of $\kappa$ used here is in line with the realistic values relevant to microfluidic experiments \citep{Pan2013,Qin2017,Qin2019}. See \cite{Morozov2022} for the estimates of $\kappa$ and its relationship to kinetic theories of dilute polymer solutions. 

The starting point of our analysis are the narwhal travelling wave solutions determined in \cite{Morozov2022} by numerically advancing the two-dimensional version of Eqs.~\eqref{eq:PTT.constitutive}-\eqref{eq:PTT.incompressibility} in time until a steady state was reached. We denote those states by $s_{2D}=\{\bm{u}_{2D}, p_{2D}, \bm{c}_{2D}\}(x-\phi_{2D} t, y)$, where $\phi_{2D}$ is the phase speed of the travelling wave. For each narwhal state, the corresponding value of $\phi_{2D}$ was determined numerically by tracking the spatial position of the largest value of ${\rm tr}(\bm{c}_{2D})$ as a function of time. In what follows, $s_{2D}$ and $\phi_{2D}$ serve as the base state for our linear stability calculations.

To probe the linear stability of the two-dimensional narwhal states $s_{2D}$ with respect to three-dimensional infinitesimal perturbations in the form $s_{3D}=\{\delta\bm{u}, \delta p, \delta \bm{c}\}(x-\phi_{2D} t, y, z, t)$, we linearise Eqs.~\eqref{eq:PTT.constitutive}-\eqref{eq:PTT.incompressibility} in 3D around a 2D travelling wave state $s_{2D}=\{\bm{u}_{2D}, p_{2D}, \bm{c}_{2D}\}(x-\phi_{2D} t, y)$. The resulting linear equations for $s_{3D}$ are stepped forward in time to determine the growth rate of the instability. The methodology employed here is similar to that used by \cite{Orszag1983}, who studied stability of Newtonian Tollmien-Schlichting waves. 
As the linearised equations of motion for $s_{3D}$ decouple the time evolution of the Fourier modes in the $z$-direction, the full dispersion relation can be obtained from a single three-dimensional simulation. To this end, we introduce an observable $a(t,k_z)$ that is based on $s_{3D}$ as follows
\begin{align}
    a(t, k_z) & =\Big\langle\widehat{\delta c}_{xx}(x-\phi_{2D} t, y, k_z, t) \widehat{\delta c}_{xx}^*(x-\phi_{2D} t, y, k_z, t)\Big\rangle_{x,y}.
\label{eq:observable}
\end{align}
Here, the hat denotes the Fourier transform along the $z$-direction, $k_z$ is the wavenumber, and $\langle \cdots \rangle_{x,y}$ denote spatial average over the $x$- and $y$-directions. While $a(t,k_z)$ can, in principle, be defined based on any component of $s_{3D}$, here we singled out $\delta c_{xx}$ as it is the largest component of the conformation tensor and is the most dynamically active one. As we will see below, for sufficiently long times, $a(t,k_z)\sim e^{\sigma(k_z) t}$, with $\sigma(k_z)$ being twice the real part of the leading eigenvalue. Below, we refer to $\sigma=\sigma(k_z)$ as the dispersion relation. 

% Numerical details
Simulations are carried out using a parallel pseudo-spectral solver with full $3/2$-dealiasing implemented in Dedalus \citep{Burns2020} on a domain $[0, L_x]\times [-1, 1]\times [0, L_z]$, where $L_x$ and $L_z$ are the dimensionless system size in the streamwise and spanwise directions, respectively. We employ no-slip boundary conditions for the velocity, $\delta\bm{u}(x-\phi_{2D} t, y=\pm 1, z, t)=0$, and natural boundary conditions for the conformational stress tensor, where $\delta\bm{c}(x-\phi_{2D} t, y=\pm 1, z, t)$ is set to the corresponding values of Eq.~\eqref{eq:PTT.constitutive} with $\kappa=0$ at each time step \citep{Liu2013}; periodic boundary conditions are used in the streamwise and spanwise directions. All fields are represented by Fourier decompositions in the periodic directions, and by a Chebyshev expansion in the $y$-direction using $N_x \times N_y \times N_z = 256 \times 1024 \times 128$ spectral modes.
The time stepping uses a four-stage, third-order implicit-explicit Runge-Kutta method %\citep{Ascher1997} 
with the time step $dt=5\times 10^{-3}$. 
Each simulation is started from an $s_{2D}$ narwhal state. To break its translational invariance in the $z$-direction, the perturbation component $\delta c_{xx}$ is initialised with a Gaussian noise of zero mean and unit variance; the magnitude of the noise is selected to be $1\times 10^{-7}$ of the maximum value of the $c_{xx}$ component of the narwhal conformation tensor.

\section{Results}

First, we analyse linear stability of the purely elastic narwhal solutions reported in \cite{Morozov2022}; see Fig.\ref{fig:narwhals}(a), for example. We set $L_x=L_z=10$, $Re=0.01$, $\epsilon = 10^{-3}$, $\kappa = 5\cdot 10^{-5}$, and vary $\beta$ and $\Wi$. For these parameters, the laminar flow is linearly stable, and the two-dimensional travelling wave solutions appear through a sub-critical bifurcation from infinity. In what follows, the lowest value of $\Wi$ for each $\beta$ considered here corresponds to the saddle-node of the sub-critical bifurcation.

\begin{figure}
  \centerline{\includegraphics[scale=1]{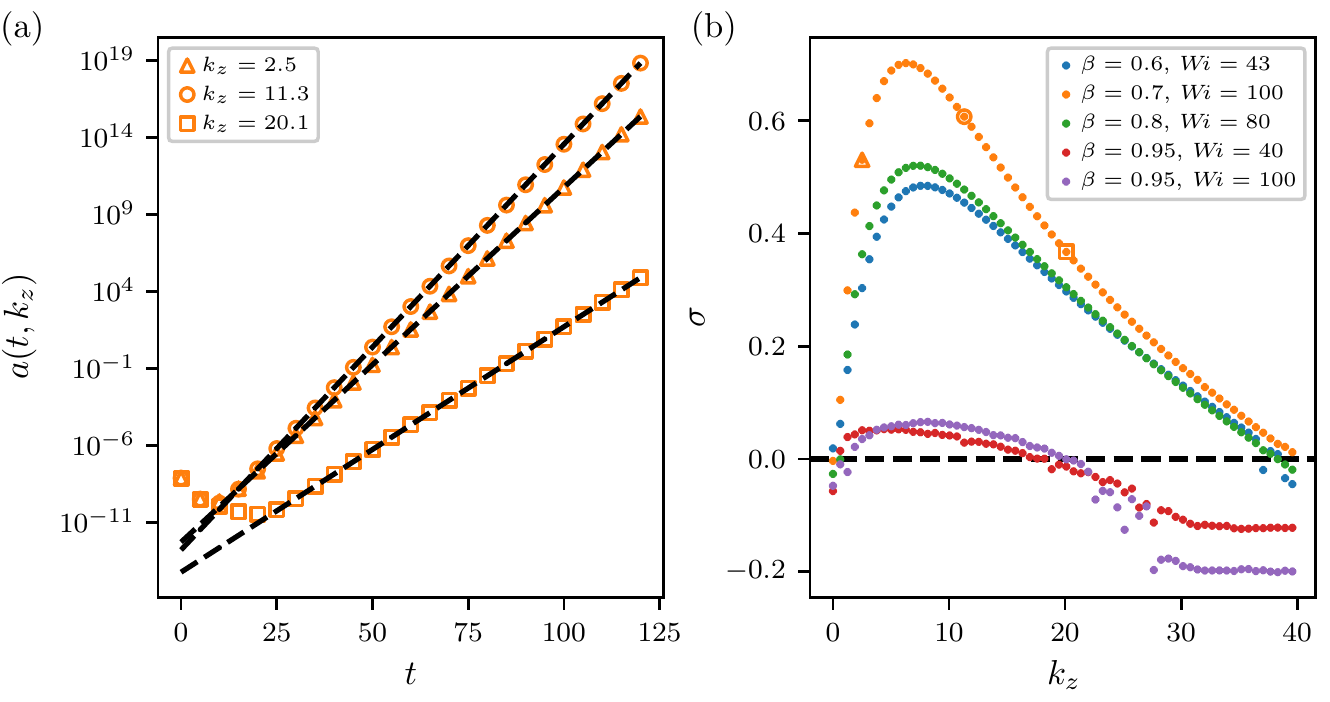}}% Images in 100% size
  \caption{(a) Time evolution of the observable $a(t, k_z)$ defined in Eq.~\eqref{eq:observable} for $(\Wi, \beta) = (100, 0.7)$ and selected wave numbers. The black dashed lines show the exponential fits used to measure the growth rate $\sigma(k_z)$. (b) Dispersion relations for representative example values of $(\Wi, \beta)$. The symbols superposed on the $(\Wi, \beta) = (100, 0.7)$ data (orange curve) represent the values of the growth rates determined from (a).}
  \label{fig:growth_rate}
\end{figure}

\begin{figure}
  \centerline{\includegraphics[width=\textwidth]{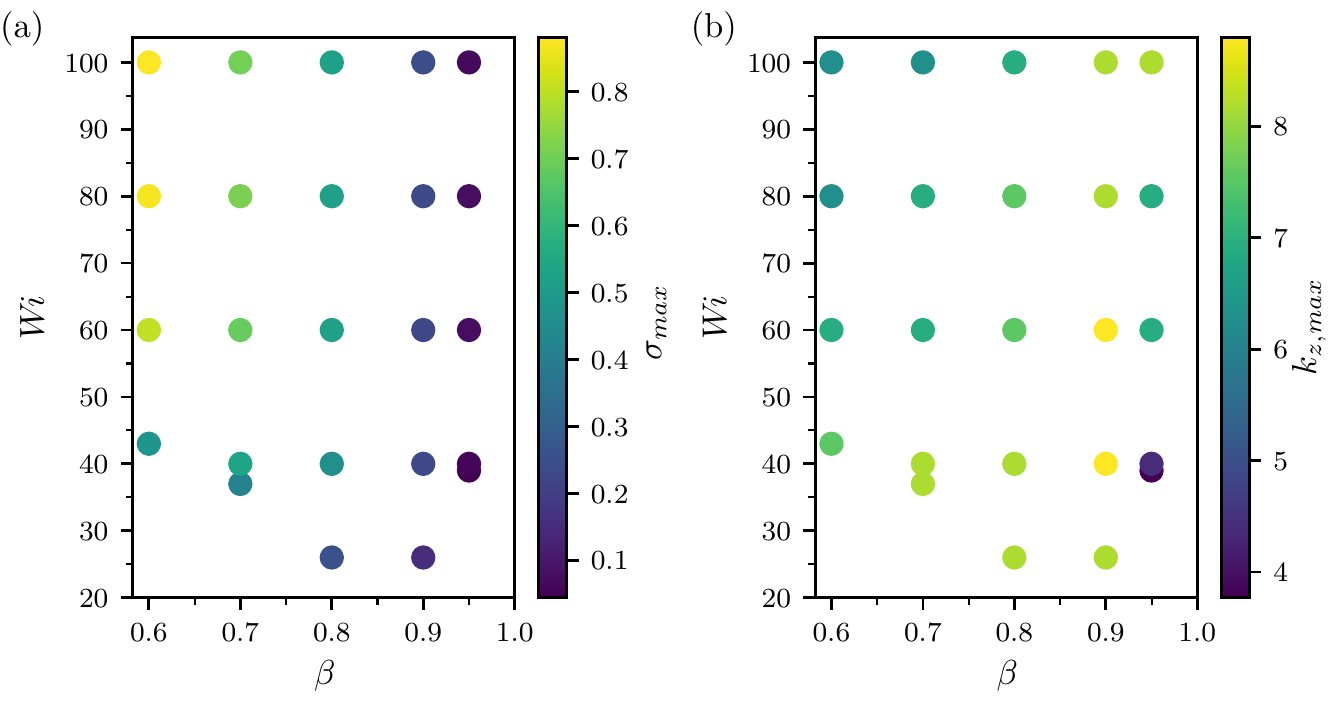}}
  \caption{Results of linear stability analysis for the purely elastic narwhal solutions found in \cite{Morozov2022}.
  (a) The growth rates of the most unstable mode, $\sigma_{max} = \max_{k_z} \sigma(k_z)$, for various $\beta$ and $\Wi$. (b) The corresponding values of $k_z$ that set the spanwise periodicity of the most unstable mode.}
  \label{fig:Wi_beta_diagram}
\end{figure}

In Fig.\ref{fig:growth_rate}(a), we present a typical evolution of the observable $a(t,k_z)$ for various values of $k_z$. As our initial condition comprises Gaussian noise, the initial projection on the most unstable eigenmode is small, and $a(t,k_z)$ initially decreases as a function of time. At sufficiently long times, the time evolution of $a(t,k_z)$ is dominated by the largest unstable eigenvalue that we determine by fitting an exponential $a(t,k_z)\sim e^{\sigma(k_z) t}$ to its long-$t$ behaviour, as discussed above. Examples of such fits are given by the dashed lines in Fig.\ref{fig:growth_rate}(a). Combining $\sigma(k_z)$ as a function of $k_z$ yields the dispersion relation for given $\beta$ and $\Wi$, and, in Fig.\ref{fig:growth_rate}(b), we show several representative examples. All dispersion relations studied in this work have the same shape, and exhibit a peak in the $\sigma(k_z)$ curves, indicating a finite-size instability. The largest growth rates and the corresponding values of $k_z$ are presented in Fig.\ref{fig:Wi_beta_diagram}. We conclude that all two-dimensional narwhal solutions become linearly unstable when embedded in a three-dimensional domain.

\begin{figure}
    \centerline{\includegraphics[width=0.9\textwidth]{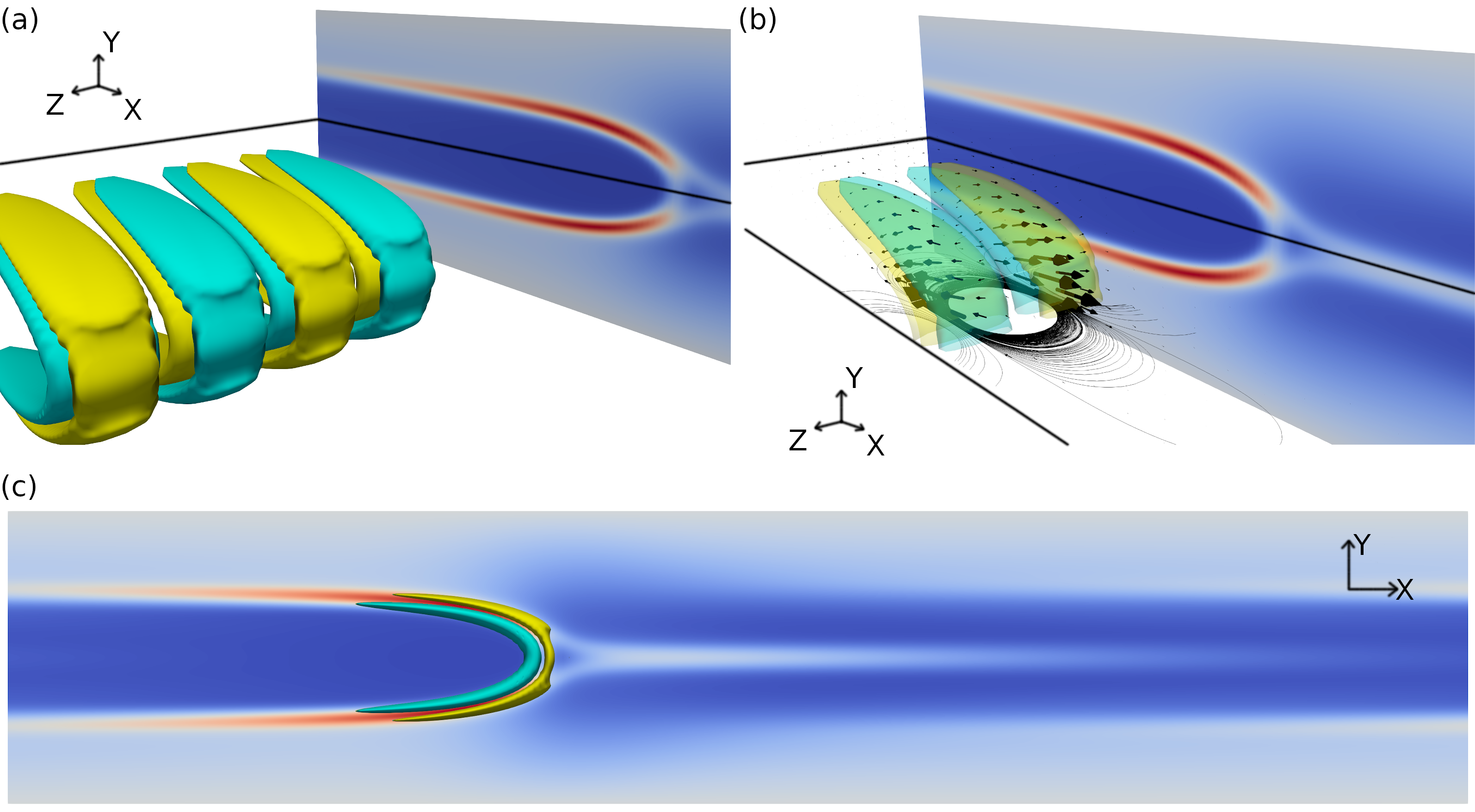}}
    \caption{
    Representative example of the three-dimensional spatial profiles of the most unstable mode for $(\Wi, \beta) = (100, 0.8)$ with $k_z\approx 6.91$. The two-dimensional base narwhal state is shown in the background of all subfigures, with the colour scheme indicating the magnitude of ${\rm tr}(\bm{c}_{2D})$. For visualisation purposes we show either one or two periods of the perturbation in the spanwise direction  only.
    (a) Isosurfaces of the perturbation ${\rm tr}(\delta \bm{c})$, with light/dark colour showing polymer extension/compression. The same structures are shown in (b) alongside the perturbation velocity field, shown by the arrows, and the centre-plane streamfunction, shown by the solid lines. 
    (c) Planar view of the base state and a half period of the perturbation that demonstrates that the perturbation stresses are localised on both sides of the narwhal `body', as discussed in the main text.}
    \label{fig:3d_figure}
\end{figure}

To gain insight into the spatial structure of the instability, in Fig.\ref{fig:3d_figure} we plot the three-dimensional profile associated with the most unstable mode (the peak of $\sigma(k_z)$) for $\beta=0.8$ and $\Wi=100$. As expected, the flow profile is periodic in the $z$-direction, and, for clarity, below we focus on a few periods in the spanwise direction only. 
As can be seen from Figs.\ref{fig:3d_figure}(a) and (c), the polymer extension associated with the perturbation, ${\rm tr} \,\delta \bm{c}$, is localised in thin sheets around the polymer extension of the base flow (the `body' of the narwhal), with regions of polymer stretch and compression alternating along the spanwise direction. This stress distribution drives a periodic array of vortices that are primarily confined to the streamwise-spanwise centre-plane, with the maxima/minima of the streamwise component of the velocity perturbation coinciding with the stretching/compression of the polymers. 

Up to this point, our analysis was restricted to the case of the sub-critical travelling-wave solutions that are disconnected from the laminar state. To show that the three-dimensional linear instability discussed above is not reliant on the $s_{2D}$ state propagating from the laminar one through a two-dimensional linear instability, we now consider two sets of parameters where this is the case. The first corresponds to the purely elastic instability identified for $1-\beta\ll1$ and $\Wi\ll1$ by \cite{Khalid2021}. To match the parameters of the linear instability identified in that work, we set $L_x=2\pi/0.75$,  $L_z=10$, $Re=0.01$, $\Wi=1700$, and $\beta = 0.997$. To demonstrate that the three-dimensional instability reported here also exists in the elasto-inertial regime, we consider the following set of parameters motivated by the work of \cite{Page2020}: $L_x=2\pi/2.18$,  $L_z=10$, $Re=90$, $\Wi=45$, and $\beta = 0.9$. In both cases we set $\epsilon = 5\cdot 10^{-5}$ and $\kappa = 3\cdot 10^{-5}$.

\begin{figure}
  \centerline{\includegraphics[width=0.60\textwidth]{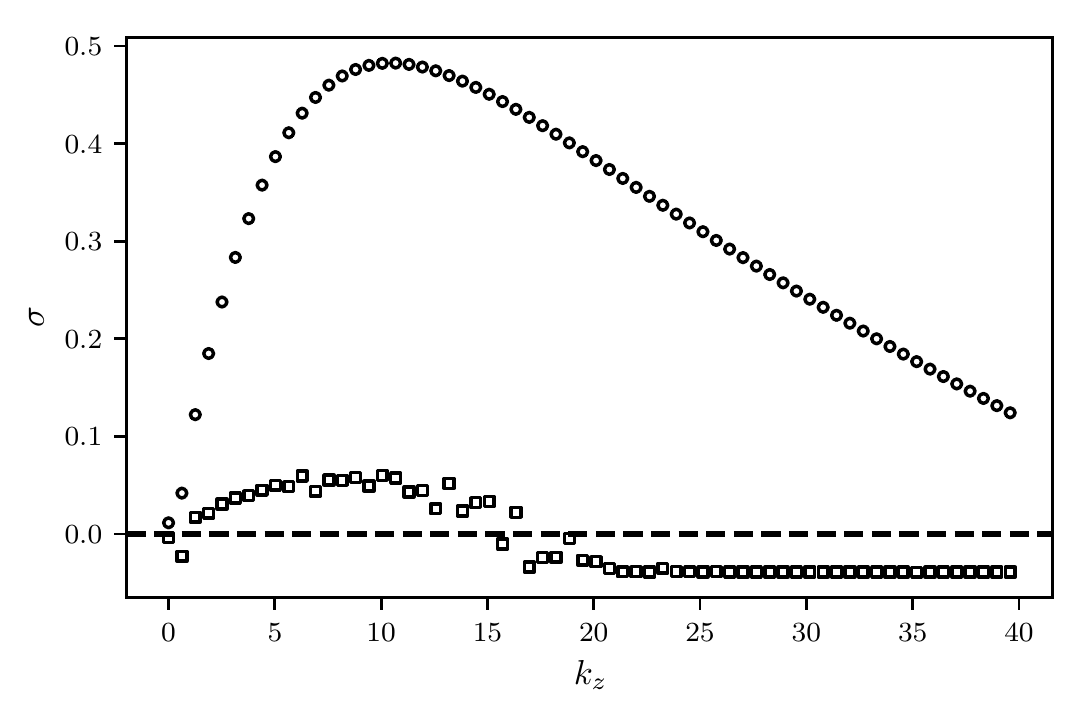}}
  \caption{%J2 is circles and Khalid is squares.
  Dispersion relations for $(\Wi, \beta, Re) = (1700,0.997,0.01)$ and $(L_x, L_z) = (2\pi/0.75, 10)$ (open squares), and $(\Wi, \beta, Re) = (45,0.9,90)$ and $(L_x, L_z) = (2\pi/2.18, 10)$ (open circles).
  }
  \label{fig:dispersion_J2_Khalid}
\end{figure}

For these two sets of parameters, we performed two-dimensional direct numerical simulations, similar to \cite{Morozov2022}, and obtained steady travelling wave solutions, shown in Fig.\ref{fig:narwhals}. Note, that our elasto-inertial narwhal solution differs slightly from the one obtained by \cite{Page2020}, as those authors considered constant-flux channel flow, while we study the constant-pressure-gradient case. In Fig.\ref{fig:dispersion_J2_Khalid}, we present the dispersion relation of the three-dimensional linear stability analysis for these two sets of parameters. Although the high-$\Wi$, high-$\beta$ case associated with the centre-mode instability in two dimensions leads to a significantly weaker three-dimensional instability, yet again, both two-dimensional structures are linearly unstable when embedded in three-dimensional domains.

\section{Discussion}

Recent numerical studies of pressure-driven channel flows of dilute polymer solutions identified narwhal travelling-wave solutions as steady coherent structures dominating purely elastic \citep{Morozov2022,buza2021} and elasto-inertial \citep{Page2020,Dubief2022} two-dimensional flows. 
In this work, we analysed linear stability of these two-dimensional structures upon their embedding in three-dimensional domains. In all parts of the parameter space $(Re,\Wi,\beta)$ studied here, both purely elastic and elasto-inertial, we find that the narwhal solutions become linearly unstable towards small perturbations periodic in the spanwise direction. The structure of the unstable mode and the shape of the dispersion relation are reminiscent of the linear instability reported in the literature in viscoelastic flows with strong spatial gradients of the polymer normal stress \citep{Renardy1988,Hinch1992} and in shear-banded flows \citep{Fielding2010,Nicolas2012}. Here, the role of the surface of large polymer extension is played by the thin sheets of high polymer stress constituting the `body' of the narwhal solution, as can be seen from Fig.\ref{fig:narwhals}. This observation is supported by Fig.\ref{fig:3d_figure}(c) that shows that the polymer stretch associated with the three-dimensional perturbation is confined to a narrow vicinity of the high-stress region of the narwhal base state. While the instability exists for a wide range of the wavenumbers $k_z$, the lengthscale associated with the most unstable mode, Fig.\ref{fig:Wi_beta_diagram}(b), is commensurate with the extent of the narwhal `body' in the wall-normal direction. As can also be seen from Fig.\ref{fig:Wi_beta_diagram}(a), the instability timescale is sufficiently short to result in a quick transition away from the base two-dimensional state. 

Stability properties of the narwhal solutions reported above mirror closely the behaviour of Tollmien-Schlichting travelling wave solutions in Newtonian pressure-driven channel flows. There, stable two-dimensional Tollmien-Schlichting waves appear through a linear instability at $Re=5772$ in sufficiently long domains \citep{Orszag1971}, and extend sub-critically to significantly lower values of $Re$ \citep{Herbert1976}. Infinitesimal three-dimensional perturbations are, however, sufficient to drive the flow away from the two-dimensional structures, leading directly towards the turbulent state \citep{Orszag1983}.
The analogy with the Tollmien-Schlichting waves has significant implications for our understanding of turbulent dynamics in the presence of polymers. Since the narwhal solutions are unstable in three dimensions, they can be seen as an effective route to ignite purely elastic or elasto-intertial turbulence. However, they are unlikely to be dynamically relevant in three dimensions, as they are too low-dimensional to organise the phase space dynamics of three-dimensional chaotic flows, as indeed are the Tollmien-Schlichting waves in the Newtonian case.
While recent studies suggested that elasto-inertial turbulence could be quasi-two-dimensional in nature \citep{Samanta2013,Dubief2013,Sid2018,Shekar2019,Shekar2021}, our results demonstrate that no conclusions about purely elastic or elasto-inertial flows can be drawn from studying strictly two-dimensional versions of such flows; their properties can only be accessed by three-dimensional simulations.

%acknowledgements
Computational resources on ARCHER2 ({\tt www.archer2.ac.uk}) were provided by the UK Turbulence Consortium ({\tt https://www.ukturbulence.co.uk/}, EPSRC grant number EP/R029326/1). We gratefully acknowledge the support we received from the Dedalus team (https://dedalus-project.org), and Keaton Burns in particular.
We acknowledge financial support from the German Academic Scholarship Foundation
(Studienstiftung des deutschen Volkes) and this work received funding from Priority Programme SPP 1881 ``Turbulent Superstructures" of the Deutsche Forschungsgemeinschaft (DFG, grant number Li3694/1). For the purpose of open access, the authors have applied a Creative Commons Attribution (CC BY) licence to any Author Accepted Manuscript version arising from this submission.

\textbf{Declaration of Interests.} The authors report no conflict of interest.

\bibliographystyle{jfm}
\bibliography{master}

\begin{thebibliography}{40}
\expandafter\ifx\csname natexlab\endcsname\relax\def\natexlab#1{#1}\fi
\def\au#1{#1} \def\ed#1{#1} \def\yr#1{#1}\def\at#1{#1}\def\jt#1{\textit{#1}}
  \def\bt#1{#1}\def\bvol#1{\textbf{#1}} \def\vol#1{#1} \def\pg#1{#1}
  \def\publ#1{#1}\def\arxiv#1{#1}\def\org#1{#1}\def\st#1{\textit{#1}}

\bibitem[{Burns} {\em et~al.\/}(2020){Burns}, {Vasil}, {Oishi}, {Lecoanet} \&
  {Brown}]{Burns2020}
{\sc \au{{Burns}, Keaton~J.}, \au{{Vasil}, Geoffrey~M.}, \au{{Oishi},
  Jeffrey~S.}, \au{{Lecoanet}, Daniel} \& \au{{Brown}, Benjamin~P.}} \yr{2020}
  \at{{Dedalus: A flexible framework for numerical simulations with spectral
  methods}}.  \jt{Phys. Rev. Res.}  \bvol{2},  \pg{023068}.

\bibitem[Buza {\em et~al.\/}(2022)Buza, Page \& Kerswell]{buza2021}
{\sc \au{Buza, Gergely}, \au{Page, Jacob} \& \au{Kerswell, Rich~R.}} \yr{2022}
  \at{Weakly nonlinear analysis of the viscoelastic instability in channel flow
  for finite and vanishing {R}eynolds numbers}.  \jt{J. Fluid Mech.}
  \bvol{940},  \pg{A11}.

\bibitem[{Castillo S\'{a}nchez} {\em et~al.\/}(2022){Castillo S\'{a}nchez},
  Jovanovi\'{c}, Kumar, Morozov, Shankar, Subramanian \& Wilson]{Sanchez2022}
{\sc \au{{Castillo S\'{a}nchez}, Hugo~A.}, \au{Jovanovi\'{c}, Mihailo~R.},
  \au{Kumar, Satish}, \au{Morozov, Alexander}, \au{Shankar, V.},
  \au{Subramanian, Ganesh} \& \au{Wilson, Helen~J.}} \yr{2022}
  \at{Understanding viscoelastic flow instabilities: Oldroyd-b and beyond}.
  \jt{J. Non-Newtonian Fluid Mech.}  \bvol{302},  \pg{104742}.

\bibitem[Chaudhary {\em et~al.\/}(2019)Chaudhary, Garg, Shankar \&
  Subramanian]{Chaudhary2019}
{\sc \au{Chaudhary, Indresh}, \au{Garg, Piyush}, \au{Shankar, V} \&
  \au{Subramanian, Ganesh}} \yr{2019}  \at{Elasto-inertial wall mode
  instabilities in viscoelastic plane {P}oiseuille flow}.  \jt{J. Fluid Mech.}
  \bvol{881},  \pg{119--163}.

\bibitem[Datta {\em et~al.\/}(2022)Datta, Ardekani, Arratia, Beris,
  Bischofberger, McKinley, Eggers, L\'opez-Aguilar, Fielding, Frishman, Graham,
  Guasto, Haward, Shen, Hormozi, Morozov, Poole, Shankar, Shaqfeh, Stark,
  Steinberg, Subramanian \& Stone]{Datta2022}
{\sc \au{Datta, Sujit~S.}, \au{Ardekani, Arezoo~M.}, \au{Arratia, Paulo~E.},
  \au{Beris, Antony~N.}, \au{Bischofberger, Irmgard}, \au{McKinley, Gareth~H.},
  \au{Eggers, Jens~G.}, \au{L\'opez-Aguilar, J.~Esteban}, \au{Fielding,
  Suzanne~M.}, \au{Frishman, Anna}, \au{Graham, Michael~D.}, \au{Guasto,
  Jeffrey~S.}, \au{Haward, Simon~J.}, \au{Shen, Amy~Q.}, \au{Hormozi, Sarah},
  \au{Morozov, Alexander}, \au{Poole, Robert~J.}, \au{Shankar, V.},
  \au{Shaqfeh, Eric S.~G.}, \au{Stark, Holger}, \au{Steinberg, Victor},
  \au{Subramanian, Ganesh} \& \au{Stone, Howard~A.}} \yr{2022}
  \at{Perspectives on viscoelastic flow instabilities and elastic turbulence}.
  \jt{Phys. Rev. Fluids}  \bvol{7},  \pg{080701}.

\bibitem[Dubief {\em et~al.\/}(2022)Dubief, Page, Kerswell, Terrapon \&
  Steinberg]{Dubief2022}
{\sc \au{Dubief, Y.}, \au{Page, J.}, \au{Kerswell, R.~R.}, \au{Terrapon, V.~E.}
  \& \au{Steinberg, V.}} \yr{2022}  \at{First coherent structure in
  elasto-inertial turbulence}.  \jt{Phys. Rev. Fluids}  \bvol{7},  \pg{073301}.

\bibitem[Dubief {\em et~al.\/}(2023)Dubief, Terrapon \& Hof]{Dubief2023}
{\sc \au{Dubief, Yves}, \au{Terrapon, Vincent~E.} \& \au{Hof, Bj\"{o}rn}}
  \yr{2023}  \at{Elasto-inertial turbulence}.  \jt{Annual Review of Fluid
  Mechanics}  \bvol{55}~(1).

\bibitem[Dubief {\em et~al.\/}(2013)Dubief, Terrapon \& Soria]{Dubief2013}
{\sc \au{Dubief, Yves}, \au{Terrapon, Vincent~E.} \& \au{Soria, Julio}}
  \yr{2013}  \at{On the mechanism of elasto-inertial turbulence}.  \jt{Phys.
  Fluids}  \bvol{25}~(11),  \pg{110817}.

\bibitem[Eckhardt(2018)]{Eckhardt2018}
{\sc \au{Eckhardt, Bruno}} \yr{2018}  \at{Transition to turbulence in shear
  flows}.  \jt{Physica A}  \bvol{504},  \pg{121--129}.

\bibitem[Fielding \& Wilson(2010)]{Fielding2010}
{\sc \au{Fielding, Suzanne~M} \& \au{Wilson, Helen~J}} \yr{2010}  \at{Shear
  banding and interfacial instability in planar {P}oiseuille flow}.  \jt{J. of
  Non-Newtonian Fluid Mech.}  \bvol{165},  \pg{196--202}.

\bibitem[Gorodtsov \& Leonov(1967)]{Gorodtsov1967}
{\sc \au{Gorodtsov, V.A.} \& \au{Leonov, A.I.}} \yr{1967}  \at{On a linear
  instability of a plane parallel {C}ouette flow of viscoelastic fluid}.
  \jt{J. Appl. Math. Mech.}  \bvol{31},  \pg{310--319}.

\bibitem[Graham \& Floryan(2021)]{Graham2021}
{\sc \au{Graham, Michael~D.} \& \au{Floryan, Daniel}} \yr{2021}  \at{Exact
  coherent states and the nonlinear dynamics of wall-bounded turbulent flows}.
  \jt{Annu. Rev. Fluid Mech.}  \bvol{53}~(1),  \pg{227--253}.

\bibitem[Herbert(1976)]{Herbert1976}
{\sc \au{Herbert, T.}} \yr{1976}  \at{Periodic secondary motions in a plane
  channel}.  \bt{In {\em Proceedings of the Fifth International Conference on
  Numerical Methods in Fluid Dynamics, June 28-July 3, 1976, Twente University
  of Technology\/} (ed. \ed{A.~I. van~de Vooren \& P.~J. Zandbergen})}.
  \publ{Springer-Verlag Berlin}.

\bibitem[{Hinch \emph{et al.}}(1992)]{Hinch1992}
{\sc \au{{Hinch \emph{et al.}}, E.~J.}} \yr{1992}  \at{The instability
  mechanism for two elastic liquids being co-extruded}.  \jt{J. of
  Non-Newtonian Fluid Mech.}  \bvol{43},  \pg{311--324}.

\bibitem[Jha \& Steinberg(2020)]{jha2020preprint}
{\sc \au{Jha, Narsing~K.} \& \au{Steinberg, Victor}} \yr{2020}  \at{Universal
  coherent structures of elastic turbulence in straight channel with
  viscoelastic fluid flow}.  \jt{arXiv:2009.12258} .

\bibitem[Jha \& Steinberg(2021)]{Jha2021}
{\sc \au{Jha, Narsing~K.} \& \au{Steinberg, Victor}} \yr{2021}  \at{Elastically
  driven {K}elvin{\textendash}{H}elmholtz-like instability in straight channel
  flow}.  \jt{Proc. Natl. Acad. Sci. U.S.A.}  \bvol{118}~(34).

\bibitem[Khalid {\em et~al.\/}(2021{\natexlab{{\em a\/}}})Khalid, Chaudhary,
  Garg, Shankar \& Subramanian]{Khalid2021a}
{\sc \au{Khalid, Mohammad}, \au{Chaudhary, Indresh}, \au{Garg, Piyush},
  \au{Shankar, V.} \& \au{Subramanian, Ganesh}} \yr{2021{\natexlab{{\em a\/}}}}
   \at{The centre-mode instability of viscoelastic plane {P}oiseuille flow}.
  \jt{J. Fluid Mech.}  \bvol{915},  \pg{A43}.

\bibitem[Khalid {\em et~al.\/}(2021{\natexlab{{\em b\/}}})Khalid, Shankar \&
  Subramanian]{Khalid2021}
{\sc \au{Khalid, Mohammad}, \au{Shankar, V.} \& \au{Subramanian, Ganesh}}
  \yr{2021{\natexlab{{\em b\/}}}}  \at{Continuous pathway between the
  elasto-inertial and elastic turbulent states in viscoelastic channel flow}.
  \jt{Phys. Rev. Lett.}  \bvol{127},  \pg{134502}.

\bibitem[Li \& Steinberg(2022)]{li2022preprint}
{\sc \au{Li, Yuke} \& \au{Steinberg, Victor}} \yr{2022} Elastic instability in
  a straight channel of viscoelastic flow without prearranged perturbations.

\bibitem[Liu \& Khomami(2013)]{Liu2013}
{\sc \au{Liu, Nansheng} \& \au{Khomami, Bamin}} \yr{2013}  \at{Elastically
  induced turbulence in {T}aylor–{C}ouette flow: direct numerical simulation
  and mechanistic insight}.  \jt{J. Fluid Mech.}  \bvol{737},  \pg{R4}.

\bibitem[Morozov(2022)]{Morozov2022}
{\sc \au{Morozov, Alexander}} \yr{2022}  \at{Coherent structures in plane
  channel flow of dilute polymer solutions with vanishing inertia}.  \jt{Phys.
  Rev. Lett.}  \bvol{129},  \pg{017801}.

\bibitem[Morozov \& van Saarloos(2019)]{Morozov2019}
{\sc \au{Morozov, Alexander} \& \au{van Saarloos, Wim}} \yr{2019}
  \at{Subcritical instabilities in plane {P}oiseuille flow of an {O}ldroyd-{B}
  fluid}.  \jt{J. Stat. Phys.}  \bvol{175},  \pg{554--577}.

\bibitem[Morozov \& van Saarloos(2005)]{Morozov2005}
{\sc \au{Morozov, Alexander~N.} \& \au{van Saarloos, Wim}} \yr{2005}
  \at{Subcritical finite-amplitude solutions for plane {C}ouette flow of
  viscoelastic fluids}.  \jt{Phys. Rev. Lett.}  \bvol{95},  \pg{024501}.

\bibitem[Morozov \& van Saarloos(2007)]{Morozov2007}
{\sc \au{Morozov, Alexander~N} \& \au{van Saarloos, Wim}} \yr{2007}  \at{An
  introductory essay on subcritical instabilities and the transition to
  turbulence in visco-elastic parallel shear flows}.  \jt{Phys. Rep.}
  \bvol{447},  \pg{112--143}.

\bibitem[Nicolas \& Morozov(2012)]{Nicolas2012}
{\sc \au{Nicolas, Alexandre} \& \au{Morozov, Alexander}} \yr{2012}
  \at{Nonaxisymmetric instability of shear-banded {T}aylor-{C}ouette flow}.
  \jt{Phys. Rev. Lett.}  \bvol{108},  \pg{088302}.

\bibitem[Orszag(1971)]{Orszag1971}
{\sc \au{Orszag, Steven~A.}} \yr{1971}  \at{Accurate solution of the
  {O}rr-{S}ommerfeld stability equation}.  \jt{J. Fluid Mech.}  \bvol{50},
  \pg{689–--703}.

\bibitem[Orszag \& Patera(1983)]{Orszag1983}
{\sc \au{Orszag, Steven~A.} \& \au{Patera, Anthony~T.}} \yr{1983}
  \at{Secondary instability of wall-bounded shear flows}.  \jt{J. Fluid Mech.}
  \bvol{128},  \pg{347–385}.

\bibitem[Page {\em et~al.\/}(2020)Page, Dubief \& Kerswell]{Page2020}
{\sc \au{Page, Jacob}, \au{Dubief, Yves} \& \au{Kerswell, Rich~R.}} \yr{2020}
  \at{Exact traveling wave solutions in viscoelastic channel flow}.  \jt{Phys.
  Rev. Lett.}  \bvol{125},  \pg{154501}.

\bibitem[Pan {\em et~al.\/}(2013)Pan, Morozov, Wagner \& Arratia]{Pan2013}
{\sc \au{Pan, L.}, \au{Morozov, A.}, \au{Wagner, C.} \& \au{Arratia, P.~E.}}
  \yr{2013}  \at{Nonlinear elastic instability in channel flows at low
  {R}eynolds numbers}.  \jt{Phys. Rev. Lett.}  \bvol{110},  \pg{174502}.

\bibitem[Phan-Thien \& Tanner(1977)]{PhanThien1977}
{\sc \au{Phan-Thien, Nhan} \& \au{Tanner, Roger~I.}} \yr{1977}  \at{A new
  constitutive equation derived from network theory}.  \jt{J. Non-Newtonian
  Fluid Mech.}  \bvol{2},  \pg{353--365}.

\bibitem[Qin \& Arratia(2017)]{Qin2017}
{\sc \au{Qin, Boyand} \& \au{Arratia, Paulo~E..}} \yr{2017}  \at{Characterizing
  elastic turbulence in channel flows at low {R}eynolds number}.  \jt{Phys.
  Rev. Fluids}  \bvol{2},  \pg{083302}.

\bibitem[Qin {\em et~al.\/}(2019)Qin, Salipante, Hudson \& Arratia]{Qin2019}
{\sc \au{Qin, Boyang}, \au{Salipante, Paul~F.}, \au{Hudson, Steven~D.} \&
  \au{Arratia, Paulo~E.}} \yr{2019}  \at{Flow resistance and structures in
  viscoelastic channel flows at low {R}e}.  \jt{Phys. Rev. Lett.}  \bvol{123},
  \pg{194501}.

\bibitem[Renardy(1988)]{Renardy1988}
{\sc \au{Renardy, Y.}} \yr{1988}  \at{Stability of the interface in two-layer
  {C}ouette flow of {U}pper {C}onvected {M}axwell {L}iquids}.  \jt{J. of
  Non-Newtonian Fluid Mech.}  \bvol{28},  \pg{99--115}.

\bibitem[Samanta {\em et~al.\/}(2013)Samanta, Dubief, Holzner, Schäfer,
  Morozov, Wagner \& Hof]{Samanta2013}
{\sc \au{Samanta, Devranjan}, \au{Dubief, Yves}, \au{Holzner, Markus},
  \au{Schäfer, Christof}, \au{Morozov, Alexander~N.}, \au{Wagner, Christian}
  \& \au{Hof, Björn}} \yr{2013}  \at{Elasto-inertial turbulence}.  \jt{Proc.
  Natl. Acad. Sci. U.S.A.}  \bvol{110}~(26),  \pg{10557–10562}.

\bibitem[Shekar {\em et~al.\/}(2021)Shekar, McMullen, McKeon \&
  Graham]{Shekar2021}
{\sc \au{Shekar, Ashwin}, \au{McMullen, Ryan~M.}, \au{McKeon, Beverley~J.} \&
  \au{Graham, Michael~D.}} \yr{2021}  \at{{T}ollmien-{S}chlichting route to
  elastoinertial turbulence in channel flow}.  \jt{Phys. Rev. Fluids}
  \bvol{6},  \pg{093301}.

\bibitem[Shekar {\em et~al.\/}(2019)Shekar, McMullen, Wang, McKeon \&
  Graham]{Shekar2019}
{\sc \au{Shekar, Ashwin}, \au{McMullen, Ryan~M.}, \au{Wang, Sung-Ning},
  \au{McKeon, Beverley~J.} \& \au{Graham, Michael~D.}} \yr{2019}
  \at{Critical-layer structures and mechanisms in elastoinertial turbulence}.
  \jt{Phys. Rev. Lett.}  \bvol{122},  \pg{124503}.

\bibitem[Shnapp \& Steinberg(2021)]{shnapp2021preprint}
{\sc \au{Shnapp, Ron} \& \au{Steinberg, Victor}} \yr{2021}  \at{Non-modal
  elastic instability and elastic waves in weakly perturbed channel flow}.
  \jt{arXiv:2106.01817} .

\bibitem[Sid {\em et~al.\/}(2018)Sid, Terrapon \& Dubief]{Sid2018}
{\sc \au{Sid, S.}, \au{Terrapon, V.~E.} \& \au{Dubief, Y.}} \yr{2018}
  \at{Two-dimensional dynamics of elasto-inertial turbulence and its role in
  polymer drag reduction}.  \jt{Phys. Rev. Fluids}  \bvol{3},  \pg{011301}.

\bibitem[Steinberg(2021)]{Steinberg2021}
{\sc \au{Steinberg, Victor}} \yr{2021}  \at{Elastic turbulence: an experimental
  view on inertialess random flow}.  \jt{Annu. Rev. Fluid Mech.}  \bvol{53},
  \pg{27--58}.

\bibitem[Wilson {\em et~al.\/}(1999)Wilson, Renardy \& Renardy]{Wilson1999}
{\sc \au{Wilson, Helen~J}, \au{Renardy, Michael} \& \au{Renardy, Yuriko}}
  \yr{1999}  \at{Structure of the spectrum in zero {R}eynolds number shear flow
  of the {UCM} and {O}ldroyd-{B} liquids}.  \jt{J. Non-Newtonian Fluid Mech.}
  \bvol{80},  \pg{251--268}.

\end{thebibliography}

\end{document}